# On-chip detection of radiation guided by dielectric-loaded plasmonic waveguides

Zhanghua Han[†,||,*], Ilya P. Radko[†], Noa Mazurski[‡], Boris Desiatov[‡], Jonas Beermann[†], Ole Albrektsen[†], Uriel Levy[‡] and Sergey I. Bozhevolnyi[†,*]

[†]Department of Technology and Innovation, University of Southern Denmark, Niels Bohrs Allé 1, Odense M DK-5230, Denmark

[‡]Department of Applied Physics, The Hebrew University of Jerusalem, Jerusalem, 91904, Israel

[||]Center for Terahertz Research, China Jiliang University, Hangzhou 310018, China

*Corresponding Authors: han@cjlu.edu.cn and seib@iti.sdi.dk

ABSTRACT: We report a novel approach for on-chip electrical detection of the radiation guided by dielectric-loaded surface plasmon polariton waveguides (DLSPPW) and DLSPPW-based components. The detection is realized by fabricating DLSPPW components on the surface of a gold (Au) pad supported by a silicon (Si) substrate supplied with aluminum pads facilitating electrical connections, with the gold pad being perforated in a specific locations below the DLSPPWs in order to allow a portion of the DLSPPW-guided radiation to leak into the Si-substrate, where it is absorbed and electrically detected. We present two-dimensional photocurrent maps obtained when the laser beam is scanning across the gold pad containing the fabricated DLSPPW components that are excited via grating couplers located at the DLSPPW tapered terminations. By comparing photocurrent signals obtained when scanning over a DLSPPW straight waveguide with those related to a DLSPPW racetrack resonator, we first determine the background signal level and then the corrected DLSPPW resonator spectral response, which is found consistent with that obtained from full wave numerical simulations. The approach developed can be extended to other plasmonic waveguide configurations and advantageously used for rapid characterization of complicated plasmonic circuits.

KEYWORDS: surface plasmon polaritons, dielectric loaded surface plasmon polariton waveguides, plasmonics, photocurrent mapping

In recent years, surface plasmon polaritons that are often called surface plasmons (SPs) for brevity have been a subject of extensive research[1], attracting a considerable attention due to the exciting capability of SPs to confine guided electromagnetic radiation far beyond the diffraction limit[2] and to concentrate it into subwavelength-sized volumes with the energy density much higher than that of the incident radiation[3,4]. These properties lay the foundation for high-density photonic integration based on SP waveguides and boost a variety of nonlinear applications. As a result, many SP-based waveguide configurations have been proposed and investigated to construct photonic devices for the manipulation of SP signals at the nanoscale[5]. Besides the unique confinement ability, another advantage of using SPs is that metal films and stripes supporting the SP propagation can also be utilized to carry electrical signals, thereby enabling one to combine photonics and electronics on the same chip into plasmonic circuitry featuring nanoscale dimensions[6]. Thus thermal control of radiation propagation in dielectric-loaded surface plasmon-polariton waveguide (DLSPPW) components has been experimentally demonstrated by electrically heating the DLSPPW gold stripes so as to change the refractive index of the adjacent dielectric material[7]. However, in the quest for the eventual monolithic integration of photonics, plasmonics, and electronics all on the same chip, it is necessary not only to control the SP propagation but also to incorporate SP sources and detectors into the same circuit to convert between photonic and electronic signals with high flexibility. Plasmonic nano-lasers from optically pumped hybrid plasmonic waveguides[8] or from electrically pumped metal-insulator-metal (MIM) waveguides, with III-V semiconductors[9] and silicon nanocrystals[10] as the gain medium, have been experimentally demonstrated. On-chip all-electrical detection of SP propagation along various plasmonic waveguides with the detector in the form of field-effect transistor[11] or semiconductor-metal-semiconductor[12] have been reported. However, the development of electrical SP sources and detectors, especially for some specific plasmonic waveguide configurations such as DLSPPWs that have progressed remarkably far towards practical applications[13], is still a work-in-progress.

In this paper, we report a novel approach for on-chip electrical detection of the radiation guided as DLSPPW modes in the near infrared region by making use of the supporting and electrically connected silicon (Si) substrate. The Si-Au interface constituting a Schottky contact has a barrier energy lower than the energy band gap of silicon, and this feature has already been utilized for the realization of Schottky SP-waveguide detector at telecom wavelengths in different configurations[14,15,16]. The use of Schottky contact for the characterization of spectral and/or spatial responses of optical nano-antennas[17,18,19,20] and detection of SP waveguide modes propagating along a metal stripe on a Si substrate[21,22] has also been reported, including the observation of enhancement of the detection efficiency due to surface roughness[23]. In our case, hot charge carriers (holes within the p-type Si substrate) generated by strong electromagnetic fields associated with the excitation of SP modes arrive at the Schottky contact with a kinetic energy exceeding the Schottky barrier height and can thereby be injected into the Si substrate. We note that although the results in this paper are obtained at near infrared wavelengths, since the Si-Au Schottky barrier height is lower than the Si bandgap, the presented plasmonic detector can operate at telecom wavelengths as long as the photon energy is higher than the Schottky barrier with the detection process known as internal photoemission (IPE).

On-chip detection of DLSPPW modes is realized via photocurrent mapping with a focused laser beam raster scanning over the fabricated structure. An aluminum (Al) pad was deposited on the Si substrate, and due to the strong diffusion of Al into Si during thermal treatment, an Ohmic contact was formed between the Al pad and the substrate. A 110-nm-thick Au pad electrically isolated from the Si substrate by a 280-nm-thick $SiO_2$ spacer layer was fabricated ~100 μm away from the Al pad. Between the Al and the Au pads and also in direct electric contact with the latter, a small planar section of 110-nm-thick Au film was deposited directly on top of the Si substrate to form a Schottky contact. This section was used as a metal support for DLSPPWs and DLSPPW-based racetrack waveguide-ring resonators (WRRs)[24] investigated in our work. These structures were made of Poly (methyl methacrylate) (PMMA) and defined using electron beam lithography. To facilitate the DLSPPW excitation with a focused laser beam, the DLSPPWs were terminated with identical tapered grating structures on both ends of all waveguides [Figure 1(a)]. One of the tapers served for the DLSPPW excitation, whereas the other one was used to monitor the transmission of the guided mode through the structure. To implement electrical detection of DLSPPW modes, a small rectangular slot (200 nm by 1000 nm) through the Au film was fabricated using lift-off after Au deposition. When fabricating the DLSPPW, the position of the slot was aligned with the end of the corresponding waveguide [cf. inset on Figure 1(a)]. When a DLSPPW mode propagates through the waveguiding structure passing over the slot, plasmonic oscillations reach the Si region through the slot via, providing carriers with sufficient energy to cross the Schottky barrier and resulting thereby in an electric current between the two contact pads. Al wires were bonded to the contact pads and connected to an external electric circuit so that the DLSPP-generated current could be measured while a raster scan of the sample with a focused laser beam was made. Note that, in the following, such a generated current is termed as a photocurrent, even though it is generated by the excited DLSPPW mode passing over and penetrating into the slot.

The fabrication process of the device is based on the usage of commercially available p-type Si substrates (resistivity: 1~10 $\Omega$ cm). Wet oxidation method was first used to produce the 280 nm thick $SiO_2$ layer on the whole wafer and the unwanted $SiO_2$ was removed by wet etching using Hydrofluoric acid. During this step, the photoresist pattern defined by laser direct writing was used as the mask for protection of the $SiO_2$ pad area. Using the same method, a 400-nm-thick Al pad was realized by Al evaporation, laser direct writing and Al wet etching. Then the whole sample was annealed in $N_2/H_2$ atmosphere at 460°C to form the Al/Si Ohmic contact. The fabrications of the finer structures between the two pads rely on electron beam lithography. With the corners of the pads serving as alignment marks, the Au film region between the two pads was defined. After a thermal evaporation of 110 nm Au and a subsequent lift-off process, the

Au film region as well as the rectangular slot in the Au film was realized. Finally, the whole sample was covered with a 300-nm layer of PMMA, and DLSPPW components with tapered gratings were fabricated in the PMMA using electron beam lithography. The width of all DLSPPWs was 320 nm, whereas the radius and the length of the straight section of the racetrack WRR were 2 $\mu$m and 1.2 $\mu$m, respectively[24]. Figure 1(a) presents a microscope image of the fabricated structure with the inset showing an enlarged image of the racetrack WRR. A cross sectional sketch of the investigated structure is shown in Figure 1(b). The heavily doped region in Si underneath the Al pad is due to the strong diffusion of Al into Si during the thermal treatment.

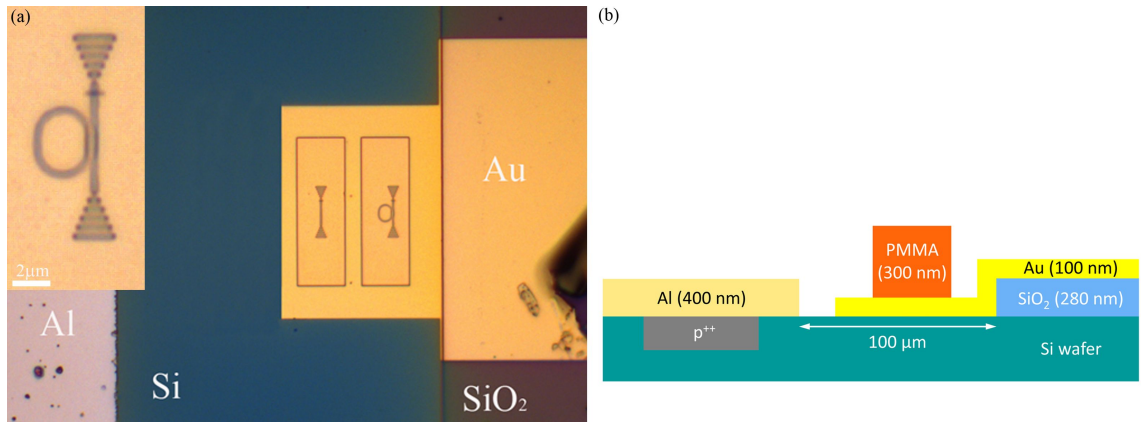

Figure 1. (a) Microscopy image taken from the top and (b) cross-sectional sketch of the investigated structure across the DLSPPW. The inset on the upper left corner of (a) presents an enlarged image of the WRR structure with a rectangular slot at the entrance of the upper part of the waveguide. Tapered gratings consisting of five ridges are optimized for DLSPPW mode excitation with a tightly focused laser beam. Al wires were bonded to Al (left) and Au (right) contact pads to measure the DLSPP-generated electric current.

Before making optical characterization, the current-voltage (I-V) characteristics of the fabricated structure was first measured using a potentiostat (Zahner, model Zennium)[25] (Figure 2). One can see that the I-V curve is consistent with the nonlinear behavior expected for a Schottky diode, demonstrating that the interface between the Au film and the Si substrate is indeed a Schottky contact. In the forward-bias condition, the current increases first slowly as a function of the applied voltage and then grows considerably when the voltage exceeds a value of around 0.5V, demonstrating that the width of the depletion region in the Schottky contact has been narrowed. In contrast, under the reverse-bias condition, the current saturates rapidly with the increased applied voltage, reaching a value of around 250 nA. This dark current value is one order of magnitude larger than that reported in the literature[14], which we attribute to the larger contact interface area used in the present structure.

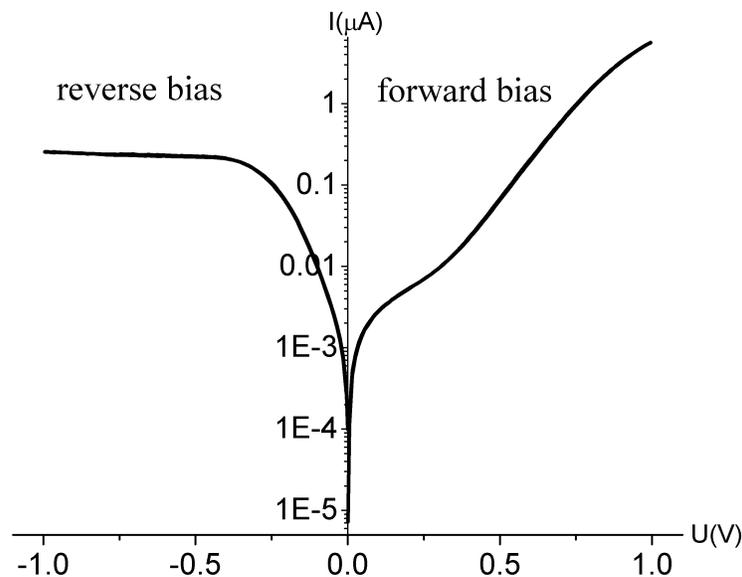

Figure 2. Measured I-V curve (semi-logarithmic scale) of the fabricated structure showing the rectifying behavior of a Schottky contact.

The racetrack WRR structure was preliminarily characterized by illuminating one of the gratings with a tunable Ti: Sapphire laser and observing the transmission through the structure with a charge-coupled device (CCD) camera (Figure 3). Though the images are influenced by scattered (at the input grating) light, different level of the device transmission could be clearly observed, manifesting itself by lighting up the slot located near the output grating, when the transmission through the racetrack WRR was at its maximum [cf. Figures. 3(a) and 3(b)].

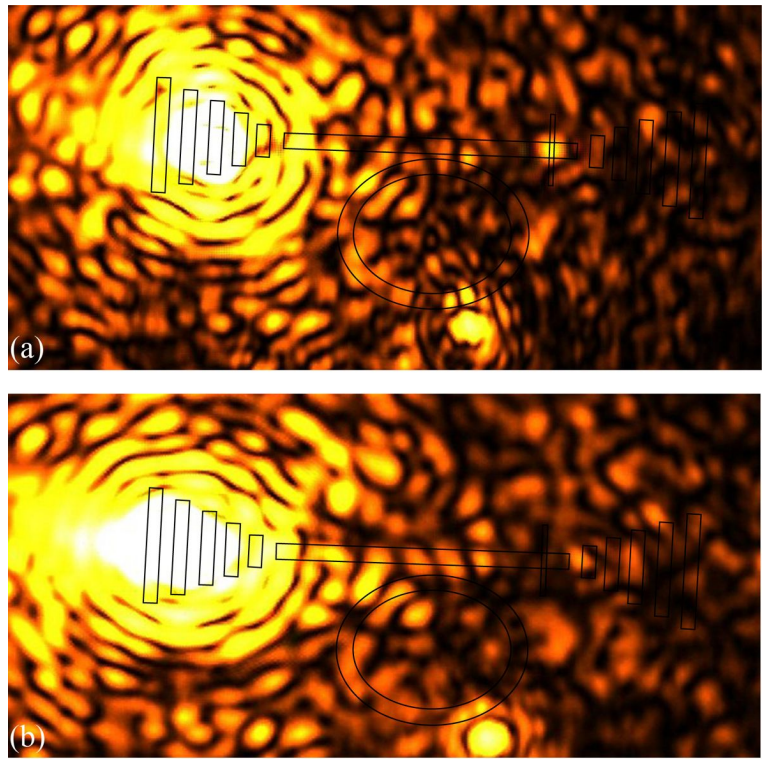

Figure 3. CCD-camera images of the whole structure when the scanning laser beam is coupled through the grating further away from the slot. The laser wavelength is (a) 750 nm and (b) 759 nm. The black lines show the profile of two gratings and the intermediate DLSPPW components as well as the rectangular slot for DLSPP detection.

For the purpose of testing the electrical response of the device, a home-made far-field scanning optical microscope[26] was used to tightly focus the Ti:Sapphire laser beam with a Mitutoyo long-working distance ×50 objective (numerical aperture 0.55) and scan the area containing the DLSPPW components. Photocurrent maps of our devices were obtained by applying a negative bias voltage of 0.1V. The polarization of the laser was set to be perpendicular to the grating ridges so that the DLSPPW mode could be efficiently excited[27]. Typical photocurrent maps of straight DLSPPW [Figure 4(a)] and DLSPPW-based racetrack WRR [Figure 4(b)], both obtained at the excitation wavelength of 750 nm, show a bright spot at the position of the rectangular slot, which stems from its direct illumination by the scanning laser beam. The photocurrent signal level at other positions of the two images is due to the excitation of a DLSPPW mode at this specific position and its further propagation to and detection through the rectangular slot. Two highlighted triangles coincide with the tapered gratings optimized for the DLSPPW mode excitation, whereas largest signals originate from inside the triangles. This is consistent with the expected coupling characteristics of gratings. A weaker photocurrent signal at the straight DLSPPW and at the racetrack WRR is observed, which might be due to the fact that the waveguide itself can work as a defect for the DLSPPW mode excitation[28], though far from being optimal given that the polarization of the laser beam is oriented along the DLSPPW. Another contribution might originate from various surface plasmon modes excited by waveguide ridges and propagating away from the DLSPPWs to be detected at the outer borders of the gold support.

The generation of the photocurrent includes the following processes. The impinge of incident laser onto the grating excites the DLSPPW mode, which propagates through the DLSPP components and reaches the subwavelength-sized slot in the Au film, where a small proportion of DLSPP mode penetrates the slot as a gap-plasmon mode that both drives hot charge carriers out of gold (through the Schottky barrier) and excites electron-hole pairs in the Si region. These charges move towards the corresponding contacts, Al and Au pads, producing the photocurrent. The generated photocurrent therefore consist, in general, of two contributions (at near-infrared wavelengths) that are very difficult to separately assess without conducting carefully designed experiments at different wavelengths. The value of the photocurrent $I$ generated by a DLSPPW mode can be determined from the following equation:

$$I = \alpha P_0 \eta T(\lambda) \qquad (1)$$

where $\alpha$ is the conversion coefficient between DLSPP power and the photocurrent, $P_0$ is the incident power from the tunable laser, $\eta$ is the coupling coefficient determined by the free space laser beam coupling to the DLSPPW mode and $T(\lambda)$ is the wavelength dependent optical transmission function (OTF) of the DLSPPW components. For example, for the racetrack WRR, the OTF will be the transmission spectrum of the WRR. Since the rectangular slot for DLSPPW mode detection is positioned near one of the DLSPPW ends, the OTF for coupling from the nearer-to-the-slot grating can be taken as unity, whereas for the opposite grating, the OTF is equal to the OTF of the DLSPPW structure. Noting that the two coupling gratings have the same design and hence equal coupling coefficients $\eta$, the OTF of the DLSPPW components can be deduced by taking the ratio of the photocurrents measured from the two opposite gratings:

$$T(\lambda) = I_f/I_n \qquad (2)$$

where indices $n$ and $f$ refer to the nearer and further gratings to the slot, respectively. To avoid uncertainty related to the exact positioning of the laser beam for mode excitation at the grating, the integrated photocurrent value from the whole grating area is used.

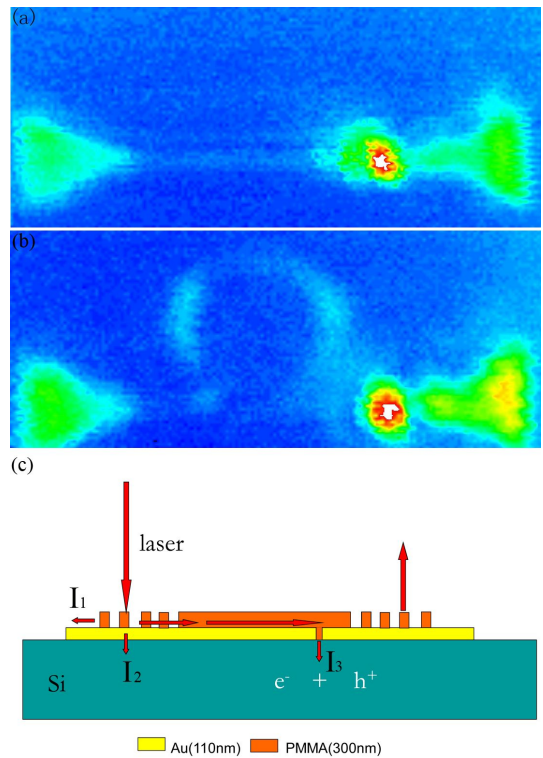

Figure 4. Typical photocurrent maps of the (a) straight DLSPPW and (b) racetrack WRR both produced at the wavelength of 750 nm; (c) Schematic showing all channels of photocurrent generation at the instant when the laser beam is scanning over one grating: (1) regular SP propagating to the gold edge, (2) light from the laser passing through the gold film, and (3) DLSPP-generated photocurrent.

Correct treatment of the experimental results requires careful accounting for noise contribution, which seemed to be rather large in our case. We assume the noise to have two primary components [Figure 4(c)]. One of the contributions comes from regular SPs excited on PMMA ridges and propagating along the flat gold–air/PMMA interface towards the edges of the Au film. This generates the strong photocurrent $I_1$ on Figure 4(c) due to a large perimeter of the Au film. The other contribution is due to direct transmission of the laser beam through the Au film, which is relatively low for a 110-nm thickness, but might still result in a considerable photocurrent $I_2$ on Figure 4(c). Together with the DLSPP-generated photocurrent $I_3$ on Figure 4(c), all three components add up to the photocurrent measured in the experiment and shown on Figures. 4(a) and 4(b). For an arbitrary DLSPPW component, the influence of the background noise can be removed by using the photocurrent values from the straight DLSPPW as a reference. Suppose $I_f^o/I_n^o$ and $I_f^-/I_n^-$ are the experimental integrated photocurrent values (cf. Eq. 2) for the racetrack WRR ("o" superscript) and the straight DLSPPW ("–" superscript) without background noise contribution being accounted for. The experimental values can then be used to evaluate the transmission spectrum of the racetrack WRR with noise contribution removed:

$$T(\lambda) = (I_f^o - I_f^-)/(I_n^o - I_n^-) \qquad (3)$$

The experimental results for OTF corrected using Eq. 3 are compared with numerical simulations based on three-dimensional (3D) full-wave finite-element method (FEM). The fitting parameter for the simulations was the length of the straight section of the racetrack WRR, which was found to be 1.21 $\mu$m versus 1.20 $\mu$m used in the fabrication design. The results agree very well in terms of free spectral range and the level of transmission at and off the resonance wavelength (Figure 5). Consequently, it is concluded that the suggested method for characterization of DLSPPW components provides accurate results, if the photocurrent background is properly taken into account. Note that although tapered gratings are used for DLSPPW mode excitation with a free-propagating laser beam, in real on-chip applications, it will be more convenient to excite DLSPPW modes by other means, e.g., by coupling a DLSPPW with a photonic waveguide[29]. In such a configuration, the influence of the background photocurrent can be highly suppressed. In this context, one can expect the photocurrent technique reported in this work to find a broad application for characterization of various plasmonic components.

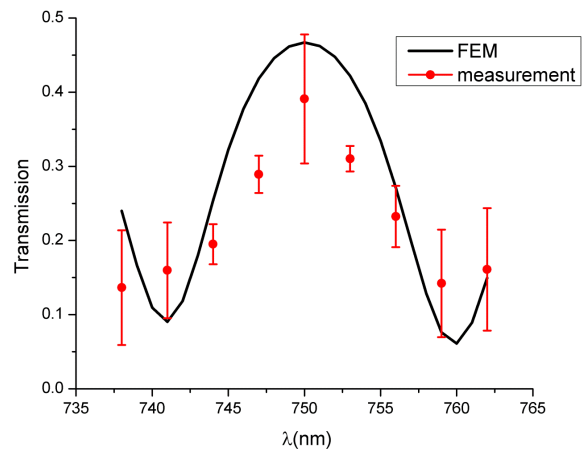

Figure. 5 Transmission spectrum of the racetrack WRR calculated using 3D FEM method (solid line) and using the experimental photocurrent values with noise subtraction using Eq. 3 (red dots).

In conclusion, we have presented a novel method for electrical detection of radiation guided by DLSPPW modes, an approach which allows on-chip integration of plasmon-based photonic and electronic circuits. By recording photocurrent maps of the sample at different illumination wavelengths, the optical transmission spectra of a DLSPPW racetrack resonator can be deduced. The issue of background noise in the racetrack measurements is dealt with by comparing the photocurrent maps of the racetrack structure with those of a straight DLSPPW. The experimental results with noise contribution being filtered out were found to agree well with full-wave 3D numerical simulations. It is noted that in real applications, one can use an opaque gold pad and a more advantageous scheme of DLSPPW mode excitation, minimizing thereby the noise contribution. The technique presented allows its further extension to more sophisticated DLSPPW-based components as well as other plasmonic waveguide configurations. In addition, the use of Al-Au contact allows the approach to be used in the telecom wavelength range as long as the photon energy is above the height of the Schottky barrier.

**Notes**

The authors declare no competing financial interest.

**Acknowledgement.**

This work was supported by the Danish Council for Independent Research (the FTP project ANAP, Contract No.09-072949) and the European Research Council, Grant 341054 (PLAQNAP). Z. Han also acknowledges the support from the National Natural Science Foundation of China (Grant No. 61107042) and the helps from Dr. Cesar E. Garcia-Ortiz on the 3D rendering.